\documentclass[pre,amsmath,onecolumn]{revtex4-2}
\usepackage[normalem]{ulem}
\usepackage{amsfonts} 
\usepackage{amsmath} 
\usepackage{color} 
\usepackage{hyperref}
\setcitestyle{square}
\usepackage{subfigure} 
\usepackage{natbib} 
\usepackage{graphicx,amssymb}

\usepackage{gensymb} 
\usepackage{mathtools}

\setcitestyle{square}

\usepackage{enumitem}

\begin{document} 

\title{Collective behavior of stock prices in the time of crisis as a response to the external stimulus}

\author{Maryam Zamani\textsuperscript{a,b,c,*},  Sander Paekivi \textsuperscript{a}, Philipp Meyer \textsuperscript{d}, Holger Kantz\textsuperscript{a}}
 \affiliation{\makebox[\textwidth][c]{a) Max-Planck Institute for the Physics of Complex Systems, Dresden D-01187, Germany}  \\{b) Max Planck Institute for the History of Science, Berlin 14195, Germany} \\{c) BIFOLD -- Berlin Institute for the Foundations of Learning and Data, Berlin 10587, Germany} \\{d) Institute of Physics \& Astronomy Karl-Liebknecht-Str. 24/25, University of Potsdam, Potsdam-Golm D-14476, Germany}}
 \email{mzamani@mpiwg-berlin.mpg.de}
%
%
\begin{abstract}
  
We analyze the interaction between stock prices of big companies in the USA and Germany using Granger Causality. 
We claim that the increase in pair-wise Granger causality interaction between prices in the times of crisis is the consequence of simultaneous response of the markets to the outside events or external stimulus that is considered as a common driver to all the stocks, not a result of real causal predictability between the prices themselves. An alternative approach through recurrence analysis in single stock price series supports this claim. The observed patterns in the price of stocks are modelled by adding a multiplicative exogenous term as the representative for external factors to the geometric Brownian motion model for stock prices. Altogether, we can detect and model the effects of the Great Recession as a consequence of the mortgage crisis in 2007/2008 as well as the impacts of the Covid out-break in early 2020.
\end{abstract}

\maketitle{}

\section{Introduction} 
\label{Introduction}

For more than a century, financial markets are described by stochastic models with completely random, memory-less fluctuations \cite{bachelier1900theorie}. The efficient market hypothesis states that current values of stocks include all information about their future. This means not only that the future performance cannot be predicted from past values of the stocks, but it is also not influenced by the present behavior of other assets. However, information flow and interdependencies between financial indices have been the topics of interest for many years \cite{fiedor2014networks,marschinski2002analysing,guo2018development}. Empirical analysis showed that financial markets behave as one entity during the time of crisis \cite{rocchi2017emerging,das2019effect}, meaning the correlation between world's financial markets increases during these times and there is a propagation of volatility between markets  \cite{junior2012correlation,sensoy2013analysis,just2020stock,das2019effect}, violating the efficient market hypothesis \cite{malkiel1989efficient,malkiel2003efficient,sewell2011history,rossi2018efficient,dias2020testing}.

The last sudden drop in stock market value happened in the early 2020, brought on by the Covid outbreak. The rapid spread of pandemic had a large impact on the markets around globe, which is uniquely different from other crisis \cite{borio2020covid,zhang2020financial}.
In this study, stock prices of 27 big companies in the USA and 11 companies in Germany since the beginning of 2000, up to the end of 2021 are analysed. We focus on the dynamics of pair-wise interaction between stocks by means of Granger causality analysis \cite{granger1969investigating} and their self-referential patterns through Auto-Recurrence Quantification Analysis (ARQA) \cite{Webber2005}. 

Specifically, we show that increases in quantifiers for predictability during the crisis, specially the Covid outbreak, is the result of the tendency of markets to behave in the same direction as a response to an extreme external driver, and not necessarily reflecting that the stock price of one company leading to another one, as could be naively interpreted from the Granger causality analysis. To demonstrate this effect, we model the stock prices using Geometrical Brownian Motion (GBM) which is an extensively used model in finance \cite{osborne1959brownian,reddy2016simulating,agustini2018stock} and add an extra term to it, representing the effect of external drivers. The GBM model assumes that the logarithm of stock values behave as Brownian motion. However, the volatility of real values of stocks in the future depends on the current value, i.e. if the current price is high, the expected price change (independent of its direction) is also high. GBM is a simple model with multiplicative noise which reproduces the price of stocks with the same mean and follows the efficient market hypothesis, however can not reproduce some observed patterns in stock prices, such as volatility clustering in the prices and volatility flow between markets during the time of crisis. Several extensions of the GBM model exist, which add features that were observed in finance data \cite{gu2012time, mao2013delay}. We show that including an external field in the GBM model, simulating an overarching influence for the market, the signals found in real data can be reproduced.  

Our paper is organized as follows; in section \ref{methods}, we explain the linear Granger causality test and the ARQA method, the tools we have used in this paper to analyse the stock prices.  In section \ref{modelling} , we introduce the Geometrical Brownian Motion model and our extension of exogenous multiplicative forcing which reproduces the observed patterns in the prices during the time of crisis. Summary and discussion come in section \ref{discussion}.

\section{Methods} 
\label{methods}
\subsection{Granger causality test between stock prices of companies}
\label{grangercausality} 

Granger causality (GC) is a statistical method, introduced for the first time in the field of econometrics \cite{granger1969investigating,granger1980testing,ajayi1998relationship}, to study the causal interaction between two or multiple time series. E.g. in the case of two time series, GC quantifies whether one series helps forecasting the other. The method is based on two main premises, namely
 (i) cause happens before its effect and (ii) observing the cause can improve the predictability of its effect. 
Consider two time series, ${x_{t}}$ and ${y_{t}}$, recorded simulataneously. The series ${x_{t}}$ Granger causes ${y_{t}}$ if ${x_{t}}$ contains some information which can be used for improving the forecast accuracy of ${y_{t}}$, alongside the past values of ${y_{t}}$ itself. Granger causality is thus a measure of predictability, allowing the evaluation of the usefulness of one time-series in forecasting another. The linear Granger causality from series $x_t$ to $y_t$ could be formulated mathematically using vector auto-regressive model (VAR) as follows:

\begin{equation}
   y_t = \alpha_{0} + \sum_{l=1}^{l=\tau^{'}} \alpha_{l} y_{t-l} + \sum_{l=1}^{l=\tau} \beta_{l} x_{t-l} + \epsilon_t,
   \label{LGC}
\end{equation}

where $\tau^{'}$ and $\tau$ are the maximum lag times, and $\epsilon_t$ is the residual of the forecast.
The maximum lags have to be chosen such that all past observations make a statistically significant contribution to the forecast.
In the above bi-variate causal model ${x_{t}}$ Granger causes ${y_{t}}$ if the coefficients $\beta_{l}$ are different from zero and statistically significant. The latter can be tested using an F-test with the null hypothesis that all the $\beta_{l}$ coefficients are zero. 

We apply this type of bivariate VAR model in a windowed framework, creating an asymmetric matrix of binary results: granger causal or not. Furthermore, note that Granger causality can only be applied for the stationary time series. Therefore, before applying it one has to make sure of stationary of the time series under study, in this case the log-return of the time series of prices. 

\begin{figure}[ht!]
\centering
\includegraphics[width=0.99 \textwidth]{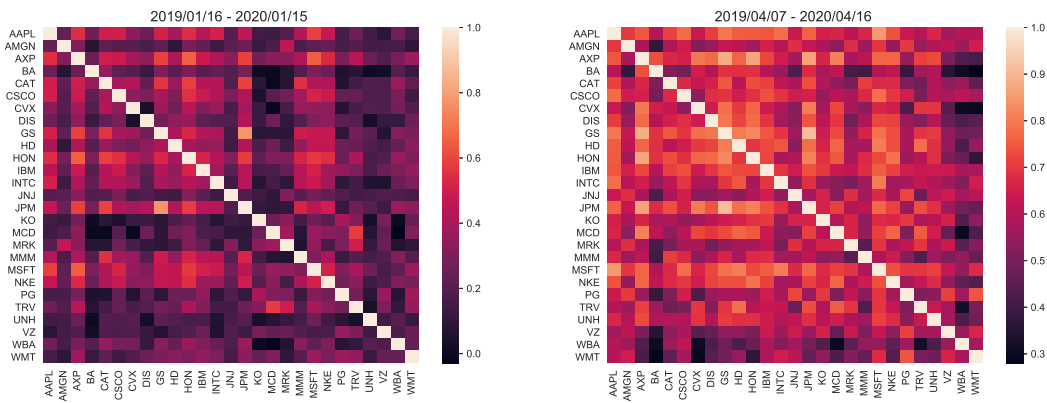}
\caption{\footnotesize{Cross-correlation matrix of 27 of big US companies during one year in two time periods. (a) before Covid outbreak (b) included the Covid crisis. Stock prices of companies show high cross-correlation during the pandemic.}}
\label{fig:Correlation_Matrix}  
\end{figure}

\begin{figure}[ht!]
\centering
\includegraphics[width=0.99 \textwidth]{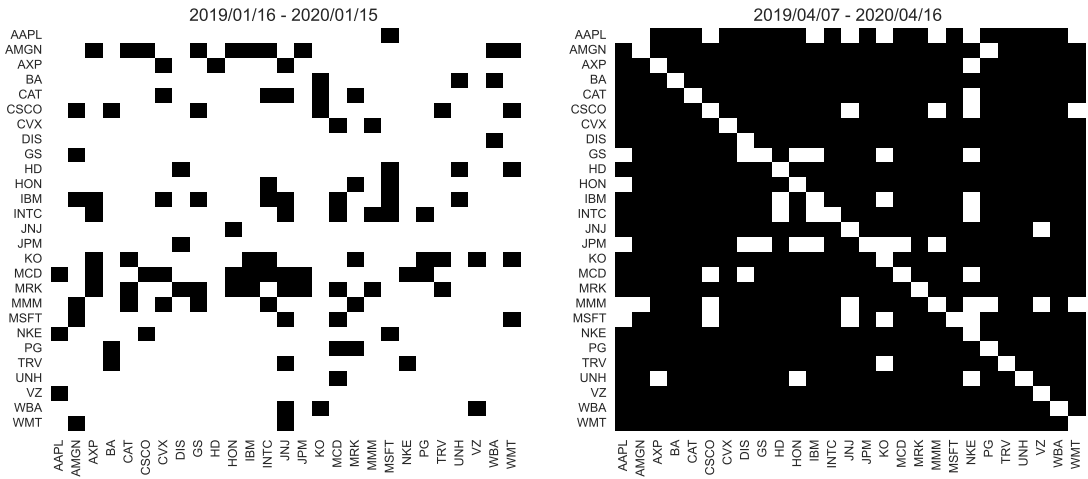}
\caption{\footnotesize{Pair-wise Granger causality matrix with $\tau=\tau'=5$ days for 27 of big US companies during, (a) 2019- before the start of pandemic in USA, (b) time window included the beginning of 2020 corresponds to the Covid outbreak. It shows the high influence of the pandemic in the interaction between companies, they behave as a one entity. Black blocks correspond to the existence of causality and white blocks refer to no causal interaction.}}
\label{fig:Causality_Matrix}
\end{figure}

We use the Granger causality test for the pairwise interaction between stock prices of companies in Dow Jones Industrial Average (DJIA) and DAX. Some companies from these two indices are excluded, because of unavailability of data during the time period of study. So altogether we analysed 27 big companies in the USA and 11 companies in Germany. Data are collected from Yahoo finance and contain the daily stock prices of the companies since the beginning of 2000. The time series of the log-returns of the prices were divided into moving windows with the length of 252 days (number of working days in a year) and the step size of 63 days. Each window is marked with the date of the middle of the window.
Figure \ref{fig:Correlation_Matrix} and \ref{fig:Causality_Matrix} show the heatmap of pairwise cross correlation and the binary 
G-causal interaction matrix (with $\tau=\tau'= 5$) between companies during two different periods before and after the outbreak of the covid pandemic in the early 2020 in the USA. It demonstrates the high cross correlation as well as causal interaction for the time window including the first 3.5 months of 2020. This interaction is more clear in the G-causality matrix in figure  \ref{fig:Causality_Matrix}(b).  

\begin{figure}[ht!]
\centering
\includegraphics[width=0.99 \textwidth]{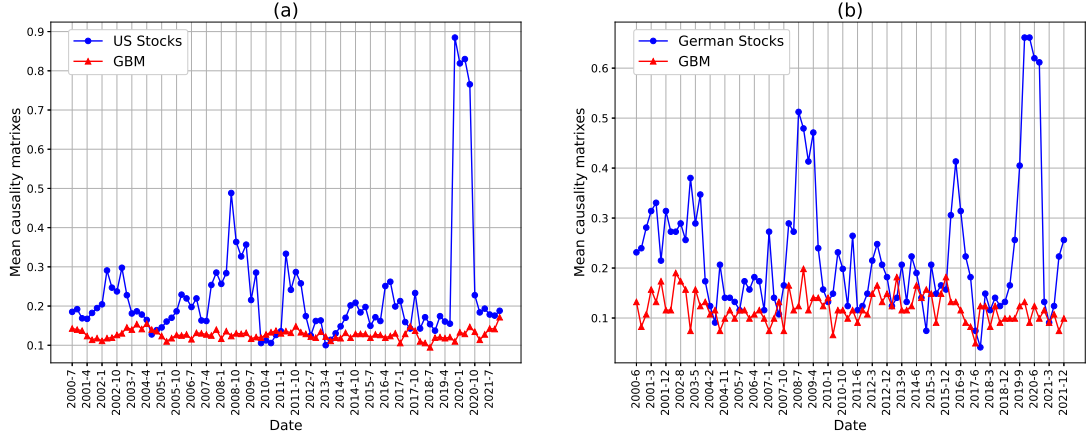}
\caption{\footnotesize{Mean of pair-wise Granger causality matrix for (a) 27 big US companies, (b) 11 big companies in Germany, and their comparison with the corresponding time series generated using Geometrical Brownian Motion.}}
\label{fig:ave_Causality}
\end{figure}

The dynamics of causal interaction during the last 21 years is captured by averaging the causal matrices of all pairs in all the moving windows as shown in figure \ref{fig:ave_Causality}. These mean values of the G-causality matrix for companies both in the States and Germany show peaks around both of the financial crisis in the period under study, namely in 2007/2008 and in 2020.
The peak in early 2020 is uniquely big and shows a massive transition towards a highly interacting system, suggested by the high level of linear interactions in the stock prices during this time. The results are compared to mean values obtained from the same number of numerically 
generated time series using Geometrical Brownian Motion (GBM). The means and variances of the GBM time series correspond to those of the original stocks. The details of the GBM model are explained in the modelling section \ref{modelling}. As shown in the figures, G-causality interaction between GBM time series shows a homogeneous behavior as expected because of similar volatility alongside the entire generated time series.

For the US companies, 80\% of the measured causality links are two standard deviations larger than the expected value from the GBM network. For the German companies, it is 54\%. The measured causality for both data sets never gets significantly lower than what would be expected from GBM. So GBM only models the unperturbed networks, while certain events (crisis) lead to a large rise in causality.

Financial crisis are not predictable, since they usually result from external factors that could not be grasped with existing models. The windowed mean causality matrix of the studied stocks however would in a naive interpretation state that during a financial crisis, predictability is at an all time high. This is due to many, if not all, of the pairwise tests yielding '1' or 'Granger causality exists', in either direction for all pairs. The latter is a common symptom of the existence of a common driver - a third influencing factor that drives both observed series, however at different rates, producing the appearance of causality between the two. In fact, given that almost the entire causality matrix is filled with '1' when covering spring 2020, it is likely that none of them reflect true causality and that all '1' caused by the existence of a common driver - global economic conditions, etc. Therefore, the Granger causality analysis in this paper yields a signal for financial crisis upon showing evidence of spurious causality - the abundance of evidence can be seen as the magnitude of the signal.

In order to verify that our hypothesis is correct, we will propose a model which contains such a common external stimulus and check how far we can reproduce the 
observed patterns in causality.
Before that, we will use Recurrence Quantification Analysis (RQA) and show that 
an equivalence of the huge peak of mean causality in early 2020 can be observed by other statistics as well.

\subsection{Recurrence quantification analysis (RQA)}
\label{RQA} 

Recurrence quantification analysis (RQA), when performed on a single time-series is often called Auto-Recurrence Quantification Analysis (ARQA), and is a non-linear method for characterizing time-series \cite{Webber2005}, robust against non-stationarities and noise on data \cite{Thiel2002, Zbilut2002, Webber2009}. The method has seen wide application in many fields (see e.g. \cite{Marwan2007}), including financial data \cite{Guhathakurta2010,Soloviev2020,Bastos2011,Aparicio2008}. 

ARQA is based on recurrence plots -- simply thresholded distance-matrices of the whole time-series. Thresholding refers to selecting a maximum distance $E$ between two values, within which the points are considered to be 'the same' or 'recurrent'. This variable is continuous for continuous time-series, discrete for discrete time-series, however for both cases the definition of a 'recurrent' value is given as $|x(t_1) - x(t_2)|<E$. A point in the recurrence plot then represents a repetition of some value at later time, and repeated sequences yield a diagonal line of the former. Analysis of the distribution of these points and especially the diagonal lines they form, can yield a variety of interpretable quantities through ARQA.

Besides the parameter $E$ defining the closeness of 'recurrence', another important component of ARQA is time delay embedding \cite{Webber2005}, which aims to reconstruct the state space dynamics of higher dimensionality, lost by considering only a scalar time-series, through supplementary delayed coordinates \cite{Takens1981,Sauer1991}. This is achieved by constructing $D$-vectors of successive elements of the time series, where 
 $D$ is called the delay and embedding. Embedded recurrence plots thus do not consider as a recurrence only a single value repeating, but a sequence, $D$ of them separated from each other by $\tau$. An embedding dimension $D>1$ may not itself be stationary however \cite{Brick2018}, as the sequence of recurrence might change its character in a non-stationary series – a feature which ARQA without embedding ignores. The issue with embedded ARQA is that it requires the additional parameter $D$ (and potentially a lag $\tau$) to be fixed. However, in applications to data which are far from representing low dimensional deterministic dynamics, the ARQA results can be rather independent of D, so that unembedded ARQA performs just as well on experimental series \cite{Iwanski1998}. Such an approach has yielded satisfactory results in previous research \cite{Arunvinthan2020, Lames2021}, and specifically in analysing financial time-series \cite{Piskun2011,Holyst2001,Kyrtsou2005,Aparicio2008, Soloviev2020}. For this reason, we also opt for employing ARQA without embedding. 

The simplest recurrence quantifier obtainable from recurrence plots is the recurrence rate ($RR$), which is simply the percentage of points that fulfill the condition $|x(t_1) - x(t_2)|<E$: 
\begin{equation}
    RR = 100 \times \frac{N_{rec}}{N_p},
\end{equation}
where $N_{rec}$ is the number of recurrent points and $N_p$ is the number of points on the recurrence plot in total. The first important quantifier is then the determinism ($DET$), calculated as:

\begin{equation}
    DET = 100 \times \frac{ N_{diag} }{ N_{rec} },
\end{equation}

where $N_{diag}$ is the number of recurrent points forming diagonal lines. It's worth pointing out that the minimal number of points required to be considered as a diagonal can vary in research, but mostly is chosen as the minimal possible, i.e. $2$. 

Another important quantifier is the laminarity $LAM$, which is analogous to $DET$ but counts the the percentage of recurrent points comprising vertical line structures rather than diagonal ones. It is calculated as:

\begin{equation}
    LAM = 100 \times \frac{ N_{vert} }{ N_{rec} },
\end{equation}

where $N_{vert}$ is the number of recurrent points forming vertical lines. In  \cite{Piskun2011}  it was demonstrated  that  laminarity  is  sensitive to critical events on markets, and in addition according to \cite{Strozzi2007} its inverse can reflect market volatility. To adequately interpret ARQA results, however, parameter selection must be understood.

As mentioned before, we use unembedded recurrence plots. Then the most relevant parameters is the recurrence defining threshold radius $E$, which can be chosen in many ways, by empirical or statistical considerations. Larger radii essentially yield more recurrent points, and can clutter the recurrence plot, making interpretation difficult or worse, making the obtained results uninformative. Among multiple suggestions a popular one is choosing $E$ that yields a low ($1-5\%$) recurrence rate while the latter scales with $E$ linearly in the log-log sense \cite{Webber2005, Wallot2018}. Another suggestion for facilitating comparisons between diverse signals is instead fixing the recurrence rate, letting $E$ vary \cite{Curtin2017, Hoorn2020}. 

We use a fixed recurrence rate and adapt $E$ in every time window and for every time series accordingly. For the given data set, 
$E$ and $RR$ scale linearly with respect to each other. However, when comparing different time-series of log-returns with fixed $E$, this leads to ARQA yielding incomparable quantifiers, reflecting mainly the change in $RR$ as a consequence of the data set's variance. Fixing $RR$ enables us to 
analyze the change in ARQA variable per-repetition, which given the efficient market hypothesis, should retain a consistent impact to ARQA. 

\begin{figure}[ht!]
\centering
\includegraphics[width=0.99 \textwidth]{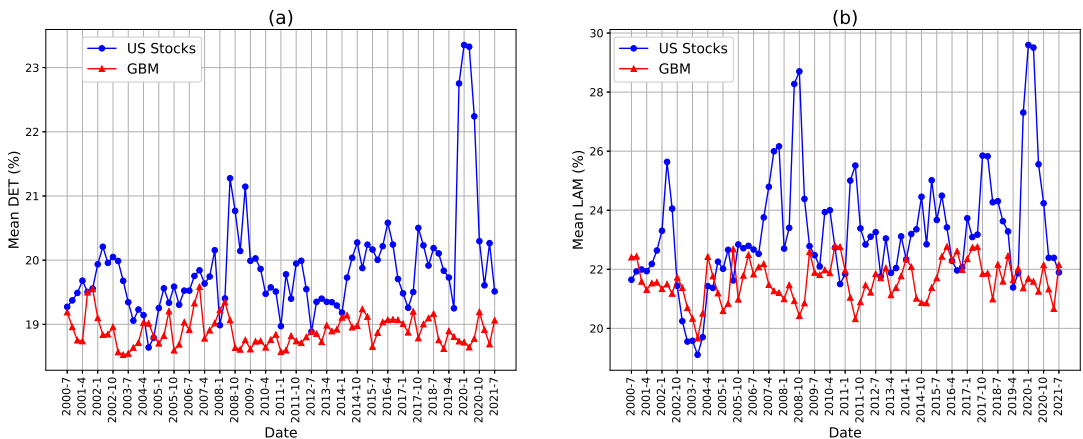}
\caption{\footnotesize{Mean of quantifiers within windowed ARQA on GBM and US stocks data, represented by the red and blue line respectively. Parameters are $RR=5\%$, no embedding. Panel (a) depicts the determinism percentage and in panel (b) the laminarity. }}
\label{fig:Stock_vs_GBM}  
\end{figure}

We carried out windowed-ARQA with the same window size and shift parameters as for the windowed Granger causality analysis, i.e. window $= 252$ and shift $= 63$ days. For each window and each stock, an auto-recurrence plot was created and ARQA carried out. Figure \ref{fig:Stock_vs_GBM} depicts $DET$ and $LAM$ across the windows, averaged between all stocks considered.

Figure \ref{fig:Stock_vs_GBM} shows peaks of DET and LAM during the two financial crisis, 2007/2008 and 2020, similarly to the Granger causality analysis results presented in Fig. \ref{fig:ave_Causality}. At first sight, this is surprising, as the employed ARQA is based on a single time-series, 
unable to quantify pairwise interactions as is done by Granger causality. This, however, is in line with our argument that the cause of these peaks is  an external common driver. The upticks in DET describe a higher frequency of repeating values within the year long window. LAM however shows similar peaks, which represent an increase of stagnancy, or a higher frequency of periods where values stay in tight bounds. These two together paint the picture of a financial crisis, as prices stagnate and oscillate around a value. 

In the next section, we introduce the GBM model in an external field, simulating the assumed abstract external influence, and discuss it in the context of Granger causality and ARQA.

\section{Modelling}
\label{modelling}

We first briefly discuss the GBM model, which is considered as a standard model for stock price dynamics \cite{mantegna1999introduction}, then we introduce a new approach by adding an external noise to the GBM model resembling the influence of external drivers that affect the dynamics of markets, which are different from simple Brownian motion or Wiener process. The GBM model can reproduce the essential market movements, but not the observed patterns in the markets especially at the time of crisis, like the one we have shown in this paper. 
GBM assumes that the logarithm of random variables ${X}$ follows a Brownian motion and can be expressed by the following stochastic differential equation \cite{fouque2000derivatives,mccauley2013stochastic},

\begin{equation}\label{stc1}
    dX(t)=\mu X(t)dt+\sigma X(t)dW(t).
\end{equation}
When modelling stock prices by GBM, $X(t)$ in above equation is considered as price at time $t$, $\mu$ is the drift term or the mean of the price and $\sigma$ is volatility. $dW$ is the increment of Wiener process which is white Gaussian noise. Using Ito's lemma, the price at time $t$ by solving above equation is

\begin{equation}
    X(t)=X_{0} e^{(\mu-\frac{\sigma^{2}}{2})t+\sigma W(t)},
\end{equation}

$X_{0}$ is the price at time $t=0$. Now, consider an additional term in above equation called $H(t)$ which represents any external driver that can affect the market price in different degrees. We rewrite equation \ref{stc1} by adding an extra term related to the external driver,

\begin{equation}\label{stc2}
    dX(t)=\mu X(t)dt+\beta h(t) X(t) dt +\sigma X(t)dW(t),
\end{equation}
where $h(t)=\frac{dH(t)}{dt}$ and the  coefficient $\beta$ represents the rate of influence which is a number between 0 and 1. The solution of above equation is

\begin{equation}
    X(t)=X_{0} e^{(\mu-\frac{\sigma^{2}}{2})t+\sigma W(t)+\beta H(t)}. 
\end{equation}

\begin{figure}[ht!]
\centering
\includegraphics[width=0.99 \textwidth]{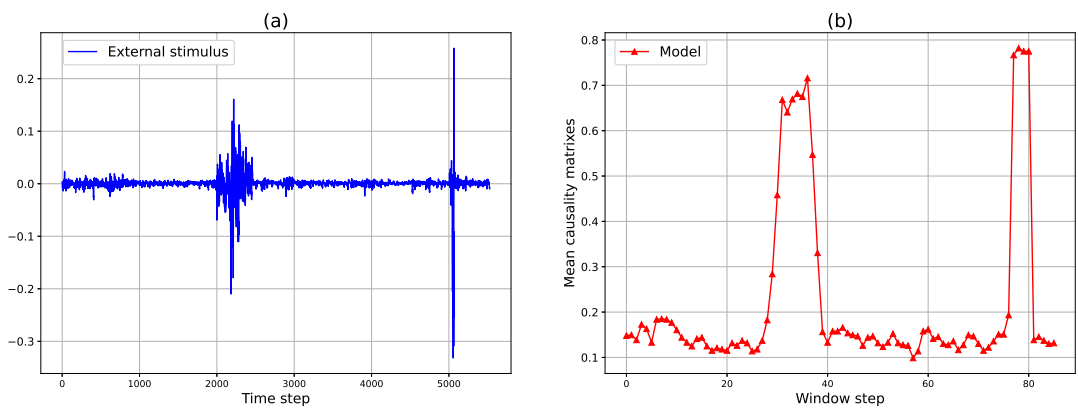}
\caption{\footnotesize{
(a) Increment $h(t)$ of external field $H$ used in the model and caculated from the log-returns of the real stock price time series. (b) Mean of causality matrices for the GBM time series in the presence of external stimulus shown in (a).}}
\label{fig:External_field_GBM}
\end{figure}

Postulating the right function for the external influence is impossible, since there are different factors that affect market movements;  effects can be short or long, weak or strong, from a tweet of some influential person over a hurricane to a terrorism attack and even bigger effects like mortgages crisis and pandemic. The fact is, we never know what is going to happen, there is an unpredictable factor with a huge influence on the market's price that can never be predicted, so that this model cannot be used for forecasting. However, retrospectively, we can construct $h(t)$ from the ensemble of observed stock price time series. 
If we assume that each single price sequence is generated by Eq.(\ref{stc2}), with individual $\mu$, $\sigma$, $\beta$, and we devide by $X(t)$, then this equation becomes an equation for the log-returns. Averaging over all stocks, 
the common driver $h(t)$ will survive, while the increments of the Wiener process $dW$ will (almost) cancel out since the ensemble mean of $dW$ equalts 0. 
What is left is the average over the unknown drift terms $\mu$, which we can remove by requiring zero mean for $h(t)$. Hence, the way to identify $h(t)$ is to calculate the ensemble average of the log-returns of the different stocks. In order to make $h(t)$ smoother, we perform an additional moving-window average.
  Figure \ref{fig:External_field_GBM}-(a) shows this signal which is used then as the time series of the increments $h(t)$ of the external stimulus $H(t)$. Evidently, there are periods during which this signal has very large amplitude, during the  first period fluctuations are larger by a factor of 
5, while during the second one by a factor of 8.

Figure \ref{fig:External_field_GBM}-(b) shows the Mean of causality matrix for 27 GBM time series in the presence of the external field $H$, whose increments are shown in figure \ref{fig:External_field_GBM}-(a).  Note that panel (a) is given in time steps, identical to the amount of work-days in the observed stocks time-series, whereas the calculated mean causality matrices are given in window-steps, which for the real stock data have been translated to dates (middle of the window). In our simulation, the coefficient $\beta$ is chosen as a random number between 0 and 1 for each generated time series. The peaks of the mean G-causality of the real data hence can be reproduced by this form of external stimulus which is added as a multiplicative exogenous term to the log-return of the prices. 

\begin{figure}[ht!]
\centering
\includegraphics[width=0.99 \textwidth]{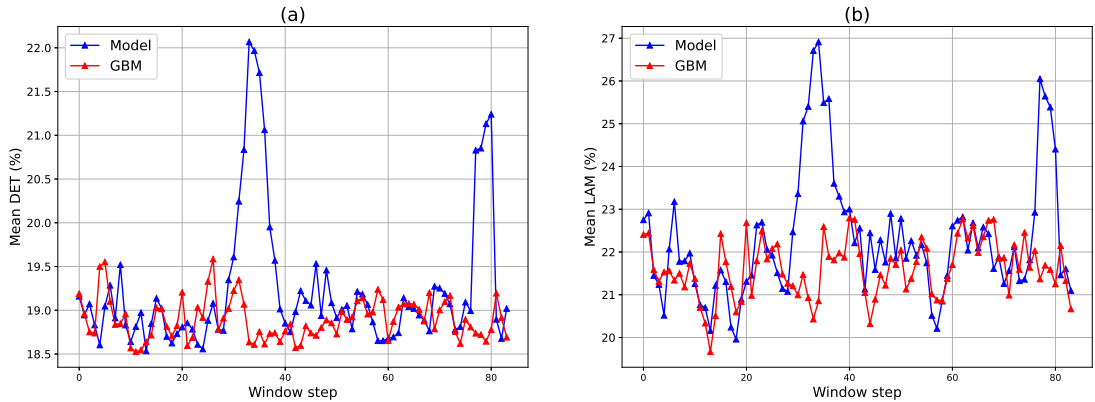}
\caption{\footnotesize{Mean of Windowed RQA quantifiers. Panel (a) is the mean of DET in the window and panel (b) is the mean of LAM in the window. A comparison between the results of the GMB trajectories (red lines) and the externally driven ones (blue lines) demonstrates signal detection by RQA for the latter.}}
\label{fig:RQA_Model_and_GBM}  
\end{figure}

Figure \ref{fig:RQA_Model_and_GBM} depicts ARQA applied to the GBM data without (Eq.(\ref{stc1})) and with an external driver (Eq.(\ref{stc2})). Both times of crisis can be seen as peaks in both DET and LAM measures, when we use the external field constructed from the real data. Considering that the GC method estimates both external forcing to be quite similar in magnitude, however different in their width, ARQA-s response to it seems to be more magnitude related, as the first (wider in time) influence causes a higher magnitude peak, albeit not by much, relatively speaking. These differences are not unexpected given the drastically different associations they operate on. What is of utmost importance to note, is that external driving can produce in both GC and ARQA analysis same kinds of signals as found in real market crisis. 

\section{Summary and Discussion} 
\label{discussion}
The interaction between time series of stock prices of 27 big US companies as well as 11 German companies during the last 21 years using the Granger causality algorithm and Recurrence analysis have been studied. 
The results from the two methods confirm each other and yield consistent results for the time evolution of stock interdependencies. These interdependencies, however, do not reflect true information exchange between the pairs of time series, but are spurious results due to the presence of an unobserved common external driver. Due to this external driver, there are periods where the efficient market hypothesis is violated, namely whenever it is strong, and this is reflected by spikes in the Granger causality and ARQA quantifiers.

These reactions have been observed during the two crisis 2007/2008 and in 2020. However, in the recent Covid crisis in the beginning of 2020, we observed uniquely strong interaction between stocks which has not been observed in other crisis. The rapid spread of the pandemic created a huge shock in markets, leading to a sudden drop of stocks' prices. Usually, there are different factors which contribute to economic crisis which are usually spontaneous and unpredictable.   We modelled the response of financial markets to these external causes by adding an exogenous term to the log-return of prices. The increase in correlation between stock prices, or the level of cross-predictability depends how extreme is the external stimulus $H(t)$. We suggested a fluctuating external field with variable intensity in time, and show a way how to construct this from past oberservations of stock prices. Simulated solutions of this model then reproduced our findings on the real data.

We consider our model as a correction to the Geometrical Brownian Motion by adding a term related to the external factors, which could be a small, short term effect like a tweet or an extreme case like pandemic, etc. These external stimuli are not produced by the market itself but have a high influence on the market movement. The simple GBM model ignores such external factors; at one hand it lacks the true prediction of prices, on the other hand it could not explain the existing observed patterns in the markets. Adding extra term in GBM model makes it possible to regenerate the observed patterns, however it does not have any predictive power since we cannot predict the external driver in the future.
The results of modelling shows the large peaks in average of Granger causality in the pairwise interactions of companies are results of the response of each individual stock to the external factor, not a real causal interaction between companies.

\bibliographystyle{aipnum4-1}
\bibliography{./bibliography2}

\end{document}